\documentclass[12pt,preprint]{aastex}

\shorttitle{The GZK cutoff revisited} \shortauthors{Booth}

\begin{document}


\title{Reassessment of the GZK cutoff in the spectrum of UHE cosmic rays in a universe with low photon-baryon ratio}

\author{Robin A.J. Booth}
\affil{Theoretical Physics, The Blackett Laboratory
\\ Imperial College, Prince Consort Road, London SW7 2BZ, UK}

\begin{abstract}
A prediction of standard Big Bang cosmology is that the observed
UHECR (ultra-high-energy cosmic rays) spectrum will exhibit a
cutoff at the GKZ limit, resulting from interaction with the
photons that constitute the cosmic microwave background. We show
that for the Quasi-Static Universe (QSU) model, in which photon
energy is an invariant in the cosmological reference frame, the
photon number density in the universe today is a factor of $10^9$
less than in the standard model. As a consequence, the mean free
path of UHECRs will exceed the horizon distance of the universe,
rendering it essentially transparent to UHECRs. The QSU model
therefore predicts that no cutoff will be observed in the UHECR
spectrum.
\end{abstract}


\section{Introduction}
Experiments to measure the flux of UHECR have detected a number of
events with energies $>10^{20} eV$ \citep{wat:2000}. These
observations present something of a puzzle. There are no obvious
mechanisms within our galaxy that could accelerate UHECRs to these
energies. It is reasonable to postulate potential extra-galactic
sources of UHECRs, e.g. quasars, but prevailing theory suggests
that these particles should never reach the Earth because of
interactions with the photons that constitute the cosmic microwave
background radiation (CMBR).  This prediction was made
independently by \citet{gre:1966} and by \citet{zat:1966}, based
on an analysis of the process $p+ \gamma \rightarrow p + \pi$
(assuming that the majority of UHECRs at these energy levels are
protons).  This showed that UHECRs with energies $> 5 \times
10^{19}$ would exceed the energy threshold for photopion
production (the GZK cutoff). Taking into account the collision
cross section for the photopion reaction, and the CMBR photon
density, the mean free path for these UHECRs would be of the order
of $10^{23}m$ (see Appendix A for analysis), implying that they
could not have originated from sources outside our galaxy.

The present generation of UHECR experiments, such as AGASA and
HiRes, are not able to generate a large enough sample from the
relatively low flux of UHECRs at these high energies
\citep{dem:2003}. The low statistical significance of the few
observations of UHECRs with energies $>10^{20} eV$ means that no
firm conclusions can be drawn at this stage. Larger experiments
are planned, such as the Pierre Auger Observatory (PAO) and EUSO,
which will increase the number of detected events by 2 orders of
magnitude.  Only then will it be possible to to prove conclusively
whether or not the predicted GZK cutoff feature is present in the
UHECR spectrum.  Nevertheless, it is reasonable to evaluate
potential cosmological scenarios that might result in the
detection of significant UHECR flux above the presumed GZK cutoff.

There have been suggestions that the observed UHECRs with energies
$>10^{20} eV$ might be due to Lorentz symmetry breaking
\citep{kif:1999,mag:2002}, or deformation of Lorentz symmetry
\citep{ame:2002}.  These predictions are based on a class of
models that postulate deviations from standard special relativity
for particles that have attained very high energies, such that
$m/R<E/E_P$, where $m$ and $E$ are respectively the particle mass
and total energy, and $E_P$ is the Planck energy. These are
generally referred to as Double Special Relativity (DSR) models.
The main criticism of this approach is that it requires either a
preferred class of inertial observers, or an ad-hoc construction
of the Lorentz deformation function to generate predicted effects
that would result in a shift of the GZK cutoff.

An alternative solution to the GZK cutoff problem is proposed in
the following sections, based on a significantly lower CMB photon
number density that arises naturally from an alternative
cosmological model \citep{boo:2002}.

\section{The Quasi-Static Universe model} The Quasi-Static Universe
(QSU)is used here as a shorthand to refer to the paradigm
described in \citet{boo:2002}.  The main feature of this model is
that the Planck scale is decoupled from the atomic scale
conventionally used as the as the basis for our measurement
system.  The changes in the dynamical behaviour of the universe
that result from this modification can not only account for most
of the problems associated with the Big Bang model, but are also
able provide a very simple explanation for the apparent
acceleration of the expansion rate of the universe that has been
observed in various high redshift supernovae studies in recent
years.  One of the principal consequences of the QSU paradigm is
that the atomic scale, as defined by the de Broglie wavelength of
sub-atomic particles, is not an appropriate reference frame for
measuring gravitational phenomena or the behaviour of photons. The
QSU model is based on the postulate that the correct reference
frame for these phenomena is in fact a cosmological frame based on
the mass and size of universe as a whole. In such a frame, photons
do not undergo any change in frequency, since as far as they are
concerned, the universe is static. Hence, it is not meaningful to
talk in terms of photons losing energy in this reference frame. In
transforming from the cosmological reference frame to our
conventional atomic frame, photons will be perceived to exhibit a
redshift as the scale factor of the universe with respect to the
atomic frame increases with time.  The crucial difference between
the QSU model and the conventional formalism is that the
relationship $E_\gamma=h\nu$ no longer holds true for photons
emitted at times $t<t_0$, where $t_0$ signifies the present time.
The energy of such 'old' photons will remain constant, at the same
value they possessed when they were first emitted.  However their
power, i.e. the energy transferred per unit time, will be reduced
in proportion to their redshift, such that $W = W_0 .{\nu
\mathord{\left/
 {\vphantom {\nu  {\nu _0 }}} \right.
 \kern-\nulldelimiterspace} {\nu _0 }}$ , where $W_0$ and $\nu_0$
are respectively the photon power and frequency at the time of
emission.  Clearly, such a modification to one the most
fundamental equations in physics will have a very significant
impact on any phenomena that involve redshifted photons, and in
particular, the CMBR.

To understand the implications of the QSU model for the CMBR, we
need to review the way in which the standard Planck black-body
distribution law is applied.  A summary of the standard derivation
of this law is provided in Appendix B. Present day observations of
the CMB give a value for the energy density of $U\simeq4\times
10^{-14}Jm^{-3}$, which from (\ref{eqn:U}) corresponds to a
temperature of $T=2.7^\circ K$.  Conventionally, the next step is
to take this temperature and use equation (\ref{eqn:RhoGamma}) to
calculate the photon number density, giving a result of $N\simeq 4
\times 10^8m^{-3}$.   However, these formulae are only valid for
black-body radiation \emph{that is in equilibrium with its
surroundings}.  It would be perfectly correct to use these
equations if we wished to deduce the photon number density for a
black-body with this temperature \emph{today}. In erroneously
applying them in the context of the relic CBMR generated by the
Big Bang, we are perpetuating the assumption inherent in going
from (\ref{eqn:dU}) to (\ref{eqn:dN2}) - that photon energy is
always equal to $h\nu$.

In the QSU model, red-shifted photons do not lose energy, so it is
not possible to determine the energy of an observed photon merely
by measuring its frequency.  It is also necessary to know the
thermal history of the photon, i.e. its frequency when it was
originally emitted.  It is this that determines the energy of the
photon.  In order to obtain the correct result for the photon
number density today that corresponds to an observed energy
density, we need to correct for the fact that the photon energy
$h\nu$ applied in going from (\ref{eqn:dU}) to (\ref{eqn:dN2})
should reflect the energy at the time of emission, or more
precisely, its energy at the time it was in black-body equilibrium
with its surroundings.

We therefore need to apply a factor of $T_{obs}/T_{equi}$ to the
expression for photon number density in (\ref{eqn:dN2}), where of
$T_{obs}$ is the observed absolute temperature of the CMBR today,
and $T_{equi}$ is the temperature at the time of black-body
equilibrium. Clearly we do not know this figure with any
precision.  However we can take an educated guess that is at least
consistent with other observational data and with plausible models
for primordial nucleosynthesis.  One such model is the neutron
decay variant of the Cold Big Bang (CBB), which predicts a maximum
reaction temperature of $\approx 10^{10} K$, with an equilibrium
temperature of the order of $\approx 10^{9} K$. Applying the
correction factor to equation (\ref{eqn:RhoGamma}) gives a
calculated photon number density of $\approx 0.3 m^{-3}$ - a
factor of $\sim 10^9$ lower than for the standard Hot Big Bang
(HBB) model. This is very close to the measured baryon number
density of the universe, giving $\eta_\gamma \simeq 1$.

\section{Implications for cosmological processes}
Having established that the value of $\eta_\gamma$ in the QSU
model is many orders of magnitude less than the conventionally
accepted value, the natural question to ask is: how does this
affect other cosmological processes?

\subsection{Primordial nucleosynthesis}
Arguably, one of the process most sensitive to changes in
$\eta_\gamma$ is that of primordial nucleosynthesis.  Applying the
conventional value of $\eta_\gamma \simeq 2 \times 10^{9}$ to the
standard model for HBB nucleosynthesis results in predicted
element abundances that are in good accord with observational
data. Relatively small changes in $\eta_\gamma$ will result in
large changes to the predicted abundances, and would therefore
appear not to be compatible with observations.  However, it has
been pointed out by \citet{agu:2001} that, provided that
appropriate changes are made to a range of initial parameters,
including $\eta_\gamma$, it is possible to construct alternative
models for primordial nucleosynthesis that will produce predicted
element abundances that are in accord with observational data. One
such model is the CBB, which takes a value of $\eta_\gamma \sim 1$
as one of its initial conditions.  It is perhaps worth mentioning
in passing that the neutron decay variant of the CBB model
predicts a photon energy of $0.78MeV$, giving a baryon to photon
energy ratio of $E_\gamma / E_B \simeq 1200$.  Under the QSU
scenario, this ratio should persist from the nucleosynthesis epoch
to the present day, and indeed, the observed value of $E_\gamma /
E_B$ is very close to this value.

\subsection{Hydrogen ionization}
\begin{figure}
  \centering
  \scalebox{1}{\includegraphics{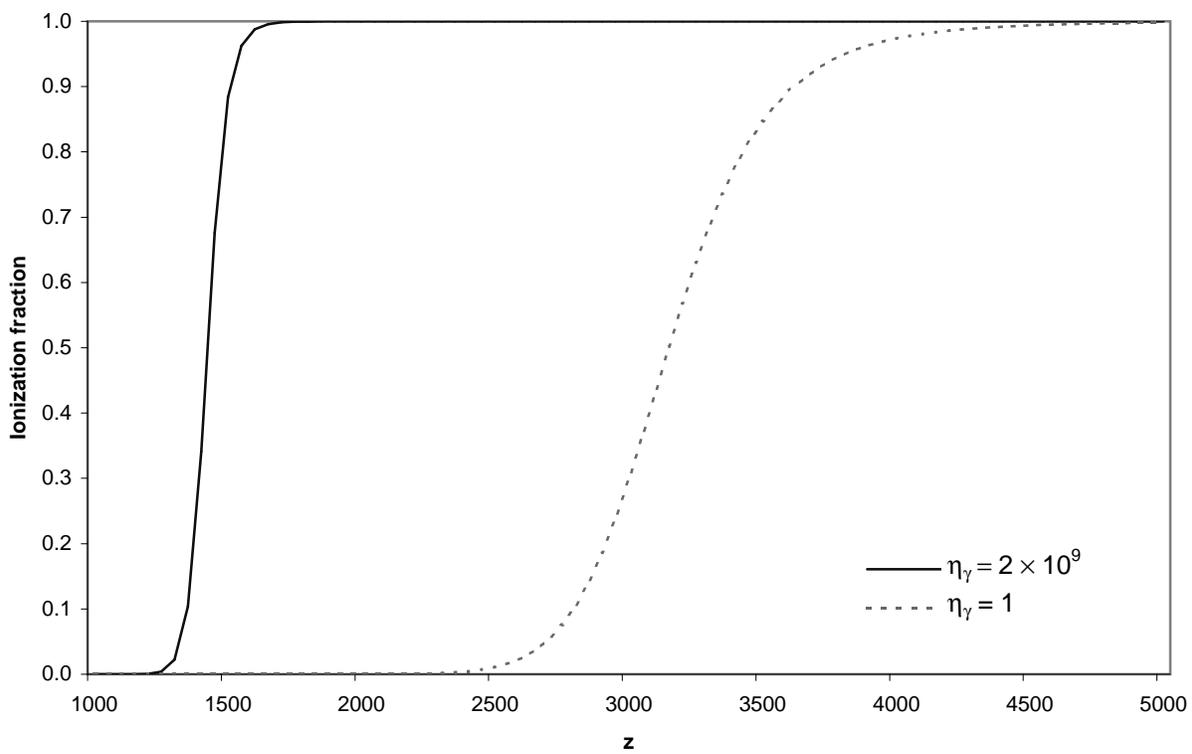}}
  \caption{Ionization fraction as a function of red-shift
  \label{fig:ionization}}
\end{figure}

Another astrophysical measurement that is linked to $\eta_\gamma$
is the hydrogen ionization fraction.  As the universe expands and
cools, protons recombine with electrons to form neutral hydrogen
when the energy of CMB photons falls below the hydrogen ionization
energy threshold. The equilibrium ionization fraction $\chi _e$ as
a function of temperature is give by the Saha equation

\begin{equation}\label{eqn:saha}
   \frac{{1 - \chi _e }}{{\chi _e }} = \frac{{4\sqrt 2 \varsigma
(3)}}{{\sqrt \pi \eta _\gamma }} \left( {\frac{T}{{m_e }}}
\right)^{\frac{3}{2}} e^{{B \mathord{\left/
 {\vphantom {B T}} \right.
 \kern-\nulldelimiterspace} T}}
\end{equation}

from which it can be see that the ionization fraction is also
dependent on $\eta_\gamma$. The ionization fraction can be
expressed as a function of redshift using the relation
$T=2.73(1+z)K$.  This is plotted in Figure~\ref{fig:ionization}
for a standard HBB cosmology with $\eta_\gamma=2 \times 10^{9}$,
and a CBB cosmology with $\eta_\gamma=1$.  This illustrates that a
reduction in the photon to baryon ratio will cause recombination
to take place at a much lower redshift, with the surface of last
scattering for the CMB occurring at $z \simeq 2400$, as compared
to $z \simeq 1100$ for the standard model.

\par At first sight it might be expected that such a change should inevitably
result in very significant modifications to the observed CMB
angular power spectrum. However, an analysis of the dynamics of
the two cosmological models reveals that the linear expansion of a
CBB universe, in comparison to the rapid initial expansion rate of
a standard model HBB universe, results in the epoch of creation of
the last scattering surface being similar, in terms of coordinate
time, for both cosmologies. Further detailed numerical analysis is
required, using a package such as CMBFAST, in order to accurately
determine the nature of the impact on the predicted CMB angular
power spectrum arising from a $\eta_\gamma=1$ cosmological model.
Nevertheless, initial indications are that such a model may well
be consistent with the observed CMB angular power spectrum.

\subsection{Reassessment of the GZK cutoff} The interaction
between CMB photons and UHECRs is dependent on three factors: the
photon energy, the capture cross-section for the photon-proton
collision, and the photon number density.  We shall now examine
how each of these factors varies in a CBB universe.

\par From (\ref{eqn:Ep}) it might be anticipated that if each CMB
photon in a QSU is a factor of $\sim 10^9$ more energetic than its
counterpart in a standard HBB universe, then the threshold energy
for the photon-proton reaction will be reduced from  $\sim 10^{20}
eV$ to  $\sim 10^{11} eV$.  However, this is unlikely to be the
case here since energy transfer from the CMBR photon to the proton
will be dependent on the photon power, rather than its total
energy. Since the photon power is, as in the standard case,
proportional to $h \nu$, there will be no change in the GZK cutoff
threshold under the QSU scenario.

\par The next factor to consider is the collision cross-section
for the photon-proton interaction.  As discussed in Appendix A,
this varies by a factor of $\sim 5$ between energies corresponding
to the main $\Delta^+$ resonance peak and higher photon energy
levels.  It is not, therefore, a significant factor in
differentiating between the two cosmological scenarios.

\par Finally, we must consider the impact of the CMB photon number
density on the mean free path of the UHECRs.  Under the QSU model,
the CMBR photon number density is a factor of $\sim 10^9$ less
than for the HBB model. From (\ref{eqn:MFP}), taking a value for
$\sigma = 100\mu \textrm{barn}$, and $n = 0.2 m^{-3}$, gives a
value for the mean free path of $\lambda = 10^{32}m$. This is
several orders of magnitude greater than the horizon distance of
the universe.  The universe is therefore effectively transparent
to UHECRs, and the QSU model predicts that there will be no
observable reduction in the flux of UHECRs as a result of the GZK
cutoff.

\section{Conclusions}
The reduced CMB photon number density predicted by the QSU model
provides a natural explanation for the observed UHECR flux with
energies $>10^{20}eV$.  However, the statistical significance of
the measurements obtained from the existing AGASA and HiRes
experiments is still to low to provide conclusive evidence for the
presence or absence of the GZK cutoff feature.  For this, we will
have to wait for the results from the next generation of UHECR
experiments, such as PAO and EUSO.

\pagebreak

\appendix

\section{The photon-proton reaction}
The principal reaction contributing to the attenuation of UHECRs
is the interaction with CMB photons
\begin{eqnarray}
    p + \gamma &\rightarrow & \Delta^+ \rightarrow p + \pi^0 \\
    p + \gamma &\rightarrow & \Delta^+ \rightarrow n + \pi^+
\end{eqnarray}
The reaction will proceed when the combined centre-of-mass energy
of the proton and photon is equal to or greater than the sum of
the pion and proton (or neutron) mass, which can be expressed as

\begin{equation}
    m_p m_\pi   + \frac{{m_\pi ^2 }}{2} \le q\left( {\sqrt {p^2  + m_p^2 }  - p\cos \theta } \right)
\end{equation}

where $q$ is the photon momentum along the $x$-axis and $p$ is the
momentum of the proton hitting the photon at an angle of $\theta$
in the $xy$ plane.  Since the pion mass is much smaller than the
proton (or neutron) mass, this expression can be simplified to

\begin{equation}\label{eqn:Ep}
    E_p  - p\cos \theta  \ge \frac{{m_p m_\pi  }}{q}
\end{equation}

For a thermal gas of relativistic bosons $\langle q\rangle \sim
2.7T$, and with $T_{CMBR}\simeq 2.7 K$, corresponding to an energy
of $2.3 \times 10^{-4} eV$, inserting the pion and proton rest
masses gives a cut-off energy of

\begin{equation}
    E_p  \sim 10^{20} eV
\end{equation}

This defines the Greisen Zatsepin Kuzmin (GZK) limit.

\par The mean free path for this reaction is given by

\begin{equation}\label{eqn:MFP}
    \lambda = \frac{1}{n \sigma}
\end{equation}

where $\sigma$ is the capture cross-section for the $p
\gamma_{CMBR}$ reaction, and $n$ is the number density of CMBR
photons.  $\sigma$ varies from a maximum of $\sim 500\mu
\textrm{barn} $ at the main $\Delta^+$ resonance to a level of
$\sim 100\mu \textrm{barn}$ at higher energy levels (see for
example \citep{muc:1999}).

\section{Black Body Radiation}

Planck's black-body distribution law can be derived in two steps.
First, by considering the number of radiation modes that can be
supported in a black-body cavity of unit volume, an expression for
the mode density can be obtained

\begin{equation}\label{eqn:dN1}
dN(\nu )  = \frac{{8\pi \nu ^2 }}{{c^3 }}d\nu
\end{equation}

The energy density as a function of frequency can then be
calculated by multiplying the mode density by the average energy
per mode.  In classical terms this would simply be $kT$, where $k$
is the Boltzmann factor and $T$ is the absolute temperature.
However, the quantum nature of photons results in the probability
distribution being skewed, so that the correct mean energy per
mode must be calculated from the Boltzmann distribution, giving
\begin{equation}\label{eqn:Ebar}
\bar E = \frac{{h\nu }}{{e^{{\raise0.5ex\hbox{$\scriptstyle {h\nu
}$} \kern-0.1em/\kern-0.15em \lower0.25ex\hbox{$\scriptstyle
{kT}$}}} - 1}}
\end{equation}

Combining (\ref{eqn:dN1} and  (\ref{eqn:Ebar}) we obtain the final
form of Planck's law for the energy density  of black-body
radiation as a function of frequency

\begin{equation}\label{eqn:dU}
dU(\nu ) = \frac{{8\pi h\nu ^3 }}{{c^3 }}.\frac{{d\nu
}}{{e^{{\raise0.5ex\hbox{$\scriptstyle {h\nu }$}
\kern-0.1em/\kern-0.15em \lower0.25ex\hbox{$\scriptstyle {kT}$}}}
- 1}}
\end{equation}

The total energy density per unit volume is then simply obtained
by integrating (\ref{eqn:dU}) to give
\begin{equation}\label{eqn:U}
U = \frac{{8\pi ^5 k^4 T^4 }}{{15c^3 h^3 }}
\end{equation}

The final expression for the photon number density as a function
of frequency is merely (\ref{eqn:dU}) divided by the photon
energy, $h\nu$

\begin{equation}\label{eqn:dN2}
dN(\nu ) = \frac{{8\pi \nu ^2 }}{{c^3 }}.\frac{{d\nu
}}{{e^{{\raise0.5ex\hbox{$\scriptstyle {h\nu }$}
\kern-0.1em/\kern-0.15em \lower0.25ex\hbox{$\scriptstyle {kT}$}}}
- 1}}
\end{equation}

The photon number density per unit volume is therefore given by

\begin{eqnarray}\label{eqn:RhoGamma}
  N & = & \int_0^\infty  {\frac{{8\pi \nu ^2 }}{{c^3 }}.\frac{{d\nu }}{{e^{{\raise0.5ex\hbox{$\scriptstyle {h\nu }$}
\kern-0.1em/\kern-0.15em \lower0.25ex\hbox{$\scriptstyle {kT}$}}}
- 1}}} \\
 &  = & \frac{{16\pi k^3 T^3 \zeta (3)}}{{c^3 h^3 }}
\end{eqnarray}

\end{document}